\documentclass[aps,prl,twocolumn,amsmath,amssymb,showpacs,nofootinbib]{revtex4}
\usepackage{graphicx,slashed,cancel,bbm,ulem}
\newcommand{\be}{\begin{equation}}
\newcommand{\ee}{\end{equation}}
\newcommand{\bea}{\begin{eqnarray}}
\newcommand{\eea}{\end{eqnarray}}

\newcommand{\sfrac}[2]{{\textstyle\frac{#1}{#2}}}
\newcommand{\D}{\mathrm{d}}
\newcommand{\E}{\mathrm{e}}
\newcommand{\I}{\mathrm{i}}

\newcommand{\Lag}{\mathcal{L}}
\newcommand{\m}{\mu}
\newcommand{\hta}{\check\tau}

\newcommand{\C}{\mathrm{c}}
\newcommand{\vs}{\vskip 5mm}
\begin{document}
\title{Hadrons on the worldline, holography, and Wilson flow}
\author{Dennis D.~Dietrich}
\affiliation{Arnold Sommerfeld Center, Ludwig-Maximilians-Universit\"at, M\"unchen, Germany}
\affiliation{Institut f\"ur Theoretische Physik, Goethe-Universit\"at, Frankfurt am Main, Germany}
\begin{abstract}
Holographic principles have impacted the way we look at strong coupling phenomena in quantum chromodynamics, strongly interacting extensions of the standard model, and {condensed-matter} physics. In real world settings, however, we still lack understanding of why and when such an approach is justified. Therefore, here, without invoking any such principle a priori, we demonstrate how such a picture arises in the worldline formulation of quantum field theory. 
Among other connections to holographic models, a warped AdS$_5$ geometry, a quantum mechanical picture, and hidden local symmetry emerge, as well as a Wilson flow (gradient flow), which extends the four-dimensional sources to five-dimensional fields and a link to the Gutzwiller trace formula. The worldline formulation also reproduces the non-relativistic case, which is important for condensed-matter physics. 
\end{abstract}
\maketitle

The AdS/CFT correspondence \cite{Maldacena:1997re} and its nonconformal and/or nonsupersymmetric generalisations have profoundly influenced physics. It impacted quantum chromodynamics (QCD) \cite{Erlich:2005qh,Karch:2006pv}, strongly interacting extensions of the Standard Model \cite{Hong:2006si,Dietrich:2008ni}, and {condensed-matter} physics \cite{Sachdev:2011wg}. It maps observables that usually have to be calculated in a strongly coupled quantum field theoretical setting onto weakly coupled quantum mechanical problems. The observed QCD hadron spectra, for example, can be approximated surprisingly well by so-called bottom-up AdS/QCD descriptions \cite{Karch:2006pv}. The bottom-up approach is still motivated by the original Maldacena conjecture, modified to accommodate necessary phenomenological features, but lacks a derivation from first principles. Therefore, it is very important to understand why and when these models represent a good approximation. In particular, they feature an extra dimension (in some approaches several), which is combined with the physical dimensions into an anti-de Sitter (AdS) space. The extended spacetime possesses conformal symmetry in addition to Lorentz invariance. Conformal symmetry is considered as the zeroth order approximation (classical massless QCD, for instance, is conformally invariant) and subsequently broken by warping or cutting off the spacetime to accommodate physical effects that break conformal symmetry such as confinement. (At least) in the ultraviolet conformal symmetry is only weakly broken in QCD, which is due to asymptotic freedom.
(For quasiconformal---e.g., technicolour \cite{walk}---theories conformality is an even better initial approximation.) Beyond encoding the conformal symmetry, the physical meaning of the extra dimension remains to be elucidated. Steps in this direction have been undertaken in light-cone holography \cite{deTeramond:2008ht}, where the extra coordinate is identified with $\zeta^2=x(1-x)\mathbf{b}_\perp^2$, where $x$ stands for the light-front momentum fraction of one of the constituents of the meson and $\mathbf{b}_\perp$ for the transverse separation of the constituents.

Here we use the worldline formulation \cite{Strassler:1992zr} of field theory and start by including two phenomenological ingredients (later, however, we will argue that the first is emergent \cite{Dietrich:2012un}): In accordance with observations (a) the squared masses of a tower of excited hadrons are spaced approximately evenly, 
and (b) in bound-state wave functions additional gluons appear to be suppressed. {Take, for instance,} large-angle hadron-hadron scattering, which is dominated by quark and not gluon exchange \cite{White:1994tj} and also \cite{Gunion:1972qi} vs \cite{Landshoff:1974ew}. Moreover, the Okubo-Zweig-Iizuka (OZI) rule \cite{Okubo:1963fa} states that processes described by Feynman diagrams that fall apart once all internal gluon lines are cut are suppressed. 

\vs

\begin{figure}[t]
\centerline{
\includegraphics[width=\columnwidth]{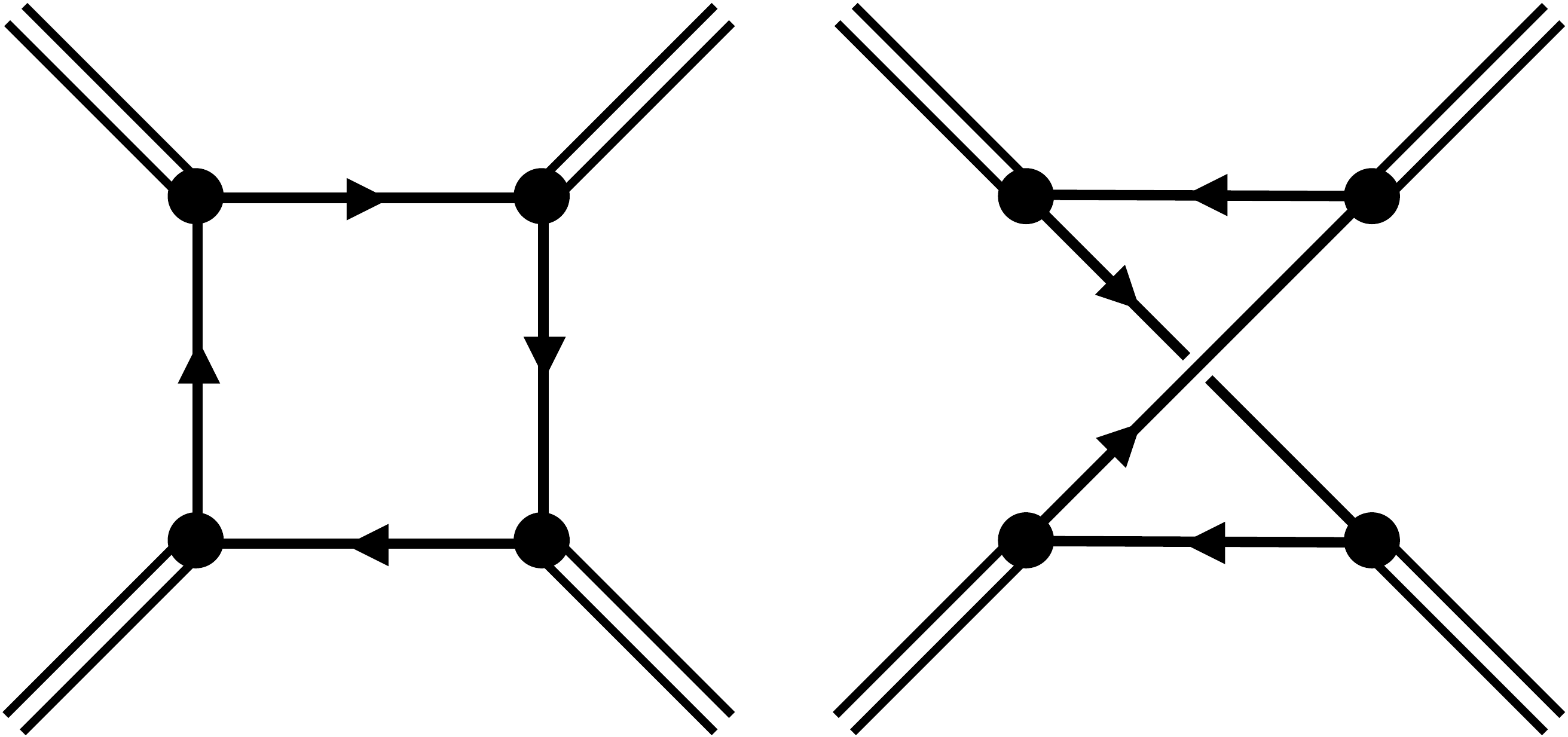}}
\caption{Some dominant diagrams. Double lines are for hadrons, single lines for quarks.}
\label{fig}
\end{figure}

Commencing with (b) the dominant diagrams {like those shown in Fig.~\ref{fig}}, i.e., without additional gluons, are included in 
\be
W=\ln\int[\D\bar\psi][\D\psi]\E^{\I\int\D^4x\bar\psi(\I\slashed D-m)\psi} .
\label{eq:genfun}
\ee 
For simplicity, we regard here only one flavour with mass $m$ and the vectorial source $V^\mu$, combined into the `covariant derivative' $D^\mu=\partial^\mu-\I V^\mu$. More sources and/or flavours can be added straightforwardly, but this does not essentially impact the following discussion. Likewise, we shall use scalar quarks below. Hence, we consider 
\be
w=-\sfrac{1}{2}\ln\mathrm{Det}(D^2+m^2)= -\sfrac{1}{2}\mathrm{Tr}\ln(D^2+m^2).
\label{eq:scalar}
\ee
In the {worldline} formalism \cite{Strassler:1992zr} this is expressed as \cite{Dietrich:2013vla_}
\begin{align}
w
=
&\int_{\varepsilon>0}^\infty\frac{\D T}{T^3}\;\E^{-m^2T}\int \D^4x_0\;\Lag,
\label{eq:ads?}
\\
\Lag
=
\frac{\mathcal{N}}{(4\pi)^2}
&\int_\mathrm{P}[\D y]\;\E^{-\int_0^T\D\tau[\frac{\dot y^2}{4}+\I \dot y \cdot V(x_0+y)]} ,
\label{eq:lag}
\end{align}
where we have continued to Euclidean space. $w$ takes the form of a Lagrangian density integrated over a soft-wall warped \cite{Karch:2006pv} AdS$_5$ space with the metric parametrisation $\D s^2=-\frac{\D T^2}{4T^2}+\frac{\D x\cdot\D x}{T}$. The extra dimension is Schwinger's proper time $T$ (with dimension length$^2$!). The corresponding integral is proper-time regulated, $T\ge\varepsilon>0$, reminiscent of the UV-brane regularisation in holographic models. The Lagrangian density $\mathcal{L}$ consists of the functional integral over all closed paths $y^\mu(0)=y^\mu(T)$ on the proper-time interval $\tau\in[0;T]$, which brings us to a quantum mechanical setting. The trajectories are shifted everywhere by the $\D^4x_0$ integration. This separation, $x^\mu=x_0^\mu+y^\mu$, is necessary to remove the zero mode of the differential operator $\partial_\tau^2$ and makes manifest translational invariance and thus energy-momentum conservation. The normalisation cancels the free path integral $\mathcal{N}\times\int[dx]\E^{-\int_0^Td\tau\frac{\dot x^2}{4}}=1$.  
The source dependent part can be rewritten as a Wilson loop $\E^{-\I\oint\D y\cdot V}$, which makes manifest the invariance under local transformations $V^\mu\rightarrow\Omega[V^\mu+\I\Omega^\dagger(\partial^\mu\Omega)]\Omega^\dagger$, i.e., emergent hidden local symmetry \cite{Bando:1984ej}.

Correlation functions of the source are obtained by carrying out the $[\D y]$ integration. 
(For plane-wave sources this is the master formula of Bern and Kosower \cite{Bern:1991aq} for $n$-point correlation functions.) 
For the two-point function, 
\begin{align}
w_2
&=
-\frac{1}{32\pi^2}\int\frac{\D^4q}{(2\pi)^4}\int^\infty_\varepsilon\frac{\D T}{T}\:{\E^{-m^2T}}\int_0^1\D\hat\tau\, \tilde l_2 ,
\nonumber\\
\tilde l_2
&=
-q^2\tilde\Pi^{\mu\nu}(q)\E^{-G_{12}q^2}\tilde V_\mu^*(q)\tilde V_\nu(q) \dot G_{12}^2 ,
\label{eq:s2}
\end{align}
where $\tilde\Pi^{\nu\lambda}(q)=\frac{q^\nu q^\lambda}{q^2}-\eta^{\nu\lambda}$, which makes transversality and hence the aforementioned local invariance manifest. (For fermionic quarks, $\dot G_{12}^2$ is replaced by $G_{12}$ up to a numerical prefactor.) Here $\tilde V_\mu(q)$ is the Fourier transform of $V_\mu(x)$. The {worldline} propagator reads $G_{ij}=G(\tau_i,\tau_j)=T\hat\tau(1-\hat\tau)$ {\cite{Strassler:1992zr}}, where 
$\tau=|\tau_i-\tau_j|$ and $\hat\tau=\tau/T$. $\dot{G}_{12}=\partial_{\tau_1}{G}_{12}$ and $\ddot{G}_{12}=\partial_{\tau_1}^2{G}_{12}$. For $\varepsilon\rightarrow 0$
\be
\int_\varepsilon^\infty\frac{dT}{T}\E^{-T[m^2+q^2(\hat\tau-\hat\tau^2)]} 
\approx
-\gamma
-\ln\{\varepsilon[m^2+q^2(\hat\tau-\hat\tau^2)]\} .
\nonumber
\ee
An imaginary part appears above the threshold $-q^2>4m^2$, signalling the decay of $V^\mu$ into its constituents. Hence, while an AdS$_5$ space and also local symmetry come about straightaway in the {worldline} formulation, an appropriate hadronic spectrum does not.

\vs

For comparison, in soft-wall AdS/QCD the two-point function is encoded in the quadratic action \cite{Karch:2006pv}
\be
S^{5D}_2=-\frac{1}{4}\int\D^4x\frac{\D T}{2T^3}\E^{-\m^2T} g^{ab}g^{cd}\mathcal{V}_{ac}\mathcal{V}_{bd} ,
\label{eq:s5}
\ee
evaluated on the classical solution. $g^{ab}$ is the inverse AdS$_5$ metric for the above parametrisation. The 4d sources are extended to 5d fields, $V_\lambda(x)\rightarrow\mathcal{V}_\lambda(x,T)$, and we fix axial gauge $\mathcal{V}_5\equiv0$. $\m$ is the warp parameter. On the saddle point only a surface term is left at $T=\varepsilon>0$, 
\be
\breve S^{5D}_2
=
\frac{1}{4}\int\frac{\D^4p}{(2\pi)^4} 
\tilde\Pi^{\nu\lambda}(p)\tilde V^*_\nu(p)\tilde V_\lambda(p)[\partial_T \tilde{\breve{v}}(p,\varepsilon)] .
\label{eq:s5sad}
\ee
The first boundary condition establishes the link between the sources in 4d to the fields in 5d, $\tilde v(p,\varepsilon)=1$, where $\tilde{\breve{\mathcal{V}}}_\lambda(p,T)=\tilde V_\lambda(p)\tilde v(p,T)$. 

{At this point, yet another tantalising parallel between the worldline and AdS/QCD expressions appears: For $\tilde v(p,\varepsilon)\equiv1$, (\ref{eq:s5}) turns into
\be
S_5\rightarrow-\frac{1}{4}\int\D^4x\frac{\D T}{2T^3}\E^{-\m^2T} g^{\mu\nu}g^{\kappa\lambda}{V}_{\mu\kappa}{V}_{\nu\lambda},
\ee
which up to a numerical factor (which can be absorbed into the normalisation of the fields or a coupling) coincides with the leading order in the so-called inverse-mass expansion \cite{Schubert:2001he} of (\ref{eq:s2}), if we identify $\mu^2\leftrightarrow m^2$.

The second boundary condition} is replaced by demanding the normalisability of the solutions. Thus,
\be
4\partial_T\tilde{\breve{v}}(p,\varepsilon)
=
-p^2[\ln(\varepsilon \m^2)+\gamma+\psi(1-\sfrac{p^2}{4\m^2})+\gamma] ,
\label{eq:dv}
\ee
where 
$\psi$ is the digamma function. 
The leading asymptotics for small $\varepsilon$ coincide with the ones above for $|p^2|\ll4\m^2$ and after identifying $\m^2\leftrightarrow m^2$ and $-p^2\leftrightarrow q^2$. The finite part of (\ref{eq:dv}) has the integral representation
\be
\gamma+\psi(1-\sfrac{p^2}{4\m^2})
=
\m^2\int_0^\infty\D T\frac{\E^{-\m^2T}-\E^{-(\m^2-\frac{p^2}{4})T}}{1-\E^{-\m^2T}} ,
\label{eq:denominator}
\ee
which can be recast into a geometric series. It reflects the presence of a tower of states with equally spaced squared masses at $q^2=4nm^2$, $n\in\mathbbm{N}$. 

\vs

Bearing this observation in mind, let us probe the response of the {worldline system} to the artificial introduction of {the} tower of states known from (\ref{eq:denominator}) into (\ref{eq:s2}) (for $m^2=0$) by a mere change of variables:
\be
\label{eq:covT}
cT=\E^{c\Theta}-1
~~\Rightarrow~~
\int_\varepsilon^\infty\frac{\D T}{T}f(T)
=
c\int_{\varepsilon}^\infty\D\Theta\frac{f[T(\Theta)]}{1-\E^{-c\Theta}}.
\ee
If $\Theta$ is to play the role of a proper-time interval, we must change the corresponding integration variable in (\ref{eq:lag}), 
\be
c\tau=\E^{c\theta}-1.
\label{eq:covtau}
\ee
This changes the form of the integrand. A standard kinetic term can be restored by ($y^\mu(0)=0=y^\mu(T)$)
\be
y_\mu=\E^{c\theta/2}\xi_\mu
\Rightarrow
\int_0^T\D\tau\Big(\frac{\D y}{\D\tau}\Big)^2
=
\int_0^\Theta\D\theta\Big[\Big(\frac{\D\xi}{\D\theta}\Big)^2+\frac{c^2}{4}\xi^2\Big].
\label{eq:coco}
\ee
Thus the system responds by cancelling the effect of the tower of states by a repulsive harmonic oscillator. (After all, the correlators of $V$ must still be the same.)
Hence, to have a physical effect one has to use a detuned setup with a different $c$ in the tower than in the harmonic oscillator, including the more minimal cases where one of the parameters is zero.

This assessment is corroborated by noting that our transformations (\ref{eq:covT}), (\ref{eq:covtau}), and (\ref{eq:coco}) coincide for $c_0=1$, $c_1=c$, and $c_2=0$ with those in \cite{de Alfaro:1976je}, 
\begin{align}
\D\Theta=\D T/U(T) ,~~~~
&\D\theta=\D\tau/U(\tau) ,~~~~
\xi^\mu=y^\mu/U^\frac{1}{2}(\tau),
\nonumber\\
U(\tau)&=c_0+c_1\tau+c_2\tau^2 .
\label{eq:covdaff}
\end{align}
which have been applied to light-front holography in \cite{Brodsky:2013ar}.
In general, the potential can have both signs $+\xi^2(c_1^2-4c_0c_2)/4$. 
Remarkably, (\ref{eq:covdaff}) introduces a scale into the Lagrangian but the action does not lose conformal invariance, as the time variable is adjusted accordingly \cite{de Alfaro:1976je}. [The $U(\tau)$ given in (\ref{eq:covdaff}) is the most general choice for which additionally the oscillator frequency is proper-time independent.] This confirms our above observation that the repulsive {harmonic oscillator} exactly compensates our substitutions and that we must detune the setup to have a net effect.

The source term is affected by the change of variables as well. We can absorb the warp factor in the source,
\be
\oint\D y\cdot V
=
\oint\D\xi\cdot\frac{\partial y}{\partial\xi}\cdot V
=
\oint\D\xi\cdot W
=
\int\D\theta\,\acute\xi\cdot W ,
\ee
where $\acute\xi=\frac{\partial\xi}{\partial\theta}$, $\frac{\partial y_\mu}{\partial\xi_\nu}=\delta_\mu^\nu\E^{c\theta/2}$, and $W\stackrel{\theta=0}{=}V$. In soft-wall AdS/QCD computations, an analogous absorption moves the warping away from the kinetic term \cite{Karch:2006pv,Dietrich:2008ni}.

\vs

The {harmonic oscillator} 
can also be understood as a two-body interaction, since
\be
\int_0^1\D\hat\tau\, [y(\tau)]^2=\frac{1}{2}\int_0^1\D\hat\tau_1\D\hat\tau_2[y(\tau_1)-y(\tau_2)]^2,
\label{eq:twobody}
\ee
with the centre-of-mass convention $\int_0^1\D\hat\tau\, y(\tau)=0$. 

An analogous potential can be obtained in a different approach to hadrons \cite{Dietrich:2012un}. It is based on the observation that, in addition to the Coulomb solution, the 
equations of motion for the gauge field admit an additional linear {equal-time} component of $A^0$ aligned with the 2 charges (of a meson for concreteness), making it consistent with translational and rotational invariance.  
Its strength is given by boundary conditions and must be tiny in QED, where we see no trace of it, but could well be present in QCD, consistent with dimensional transmutation. It has the lowest order in the coupling constant of all possible contributions. Thus, we can employ it to calculate a hadronic `Born term.' If we replace the vector source $V$ in (\ref{eq:lag}) by a gauge field $A$ and average over its action,
\be
\langle\E^{-\I\int_0^T\D\tau\,\dot y \cdot A}\rangle
=
\langle 1\rangle\:
\E^{\frac{\Lambda}{2}\int_0^T\D\tau_1\D\tau_2\,\delta(y_1^0-y_2^0)\dot y_1^0|\mathbf{y}_1-\mathbf{y}_2|\dot y_2^0},
\nonumber
\ee
where we retained only the leading {contribution coming from the linear potential}.
Doing so corresponds to adding all gauge boson exchanges
but neither matter loops (quenched, like the lowest order in \cite{'tHooft:1973jz}, which can be improved systematically \cite{'tHooft:1973jz,Armoni:2008jy}) nor gauge loops. 
Only periodic paths contribute to the path integral, and for trajectories that curve back in time exactly once (Fock states without additional pairs),
\begin{align}
\langle\E^{-\I\int_0^T\D\tau\,\dot y \cdot A}\rangle
\supset
\langle 1\rangle\:\E^{\frac{\Lambda}{2}\int_0^T\D\tau\,\mathrm{sgn}(\dot{\bar{y}}^0)\dot y^0|\mathbf{y}-\bar{\mathbf{y}}|}.
\label{eq:area}
\end{align}
Here $\bar{\mathbf{y}}(\tau)=\mathbf{y}[\bar\tau(\tau)]$, where $y^0(\bar\tau)=y^0(\tau)$. Completing the square gives rise to
\be
(\dot y^0)^2{-}2\Lambda\,\mathrm{sgn}(\dot{\bar{y}}^0)\dot y^0|\mathbf{y}-\bar{\mathbf{y}}|
=
(\dot z^0)^2-\Lambda^2(\mathbf{y}-\bar{\mathbf{y}})^2 ,
\ee
in analogy to (\ref{eq:twobody}). Here $\dot z^0=\dot y^0{-}\Lambda\,\mathrm{sgn}(\dot{\bar{y}}^0)|\mathbf{y}-\bar{\mathbf{y}}|$.
Importantly, this approach also features a linear bound-state spectrum \cite{Dietrich:2012un} without having to invoke (a).
[Additionally, in two dimensions the exponent of (\ref{eq:area}) is $\Lambda$ times the absolute area of the then planar loop. For an oriented area, this would resemble a system in a constant external magnetic field \cite{Cornwall:2003mt}, which also features linearly spaced (Landau) levels. Landau levels are related to an effective harmonic oscillator potential \cite{Greiner:1992bv}.]

Let us pause for a moment to mention that holographic pictures can be reached through large-$N_c$ arguments \cite{'tHooft:1973jz} and at least in two dimensions (\ref{eq:area}) could be reached by way of such a reasoning. The corresponding expansion parameter in real-life QCD, however, is $N_f/N_c$, which is of order 1. Therefore, as in \cite{deTeramond:2008ht,Dietrich:2012un,Dietrich:2013vla}, this approach is not followed here, but, again as in \cite{deTeramond:2008ht,Dietrich:2012un,Dietrich:2013vla}, we argue that transverse gluons are suppressed relative to instantaneous interactions. This notion is supported by observations, e.g., in large-angle scattering \cite{White:1994tj}, the resolution \cite{Gunion:1972qi} of the Landshoff paradox \cite{Landshoff:1974ew}, or the OZI sum rule \cite{Okubo:1963fa}. All these phenomena can be explained by a conditional survival of perturbation theory \cite{deTeramond:2008ht,Dietrich:2012un}, for which exist several additional indications \cite{Dokshitzer:1998qp}.

\vs

Let us continue by trying to gain more understanding about how the 4d fields are extended to the fifth dimension. To this end, let us reconsider (\ref{eq:s2}) after an integration by parts,
\be
w_2
=
\frac{-1}{{32}\pi^2}\int\frac{\D^4q}{(2\pi)^4}\int^\infty_\varepsilon\frac{\D T}{T^2}\int_0^T\D\tau{\ddot G}\tilde V_\mu^*(q)\E^{-G
q^2}\tilde V^\mu_\perp(q) ,
\nonumber
\ee
where $\tilde V^\mu_\perp=\tilde\Pi^{\mu\nu}\tilde V_\nu$. $\tilde V^\mu_\perp(q,G)=\E^{-Gq^2}\tilde V^\mu_\perp(q)$ solves
\be
(\partial_{G}+q^2)\tilde V^\mu_\perp(q,G)=0,
\ee
for
$\tilde V^\mu_\perp(q,0)=\tilde V^\mu_\perp(q)$. In position space, 
\be
V^\mu_\perp(x,G)=\E^{G\Box}V^\mu_\perp(x)=\int\frac{\D^4 x^\prime}{(4\pi G)^2}\,\E^{-\frac{(x-x^\prime)^2}{4G}}V^\mu_\perp(x^\prime);
\nonumber
\ee
i.e., the source $V(x,0)$ is smeared (smoothened) by a Gaussian whose width depends on the separation in the fifth dimension. Indeed  
\be
(\partial_{G}-\Box)V^\mu_\perp(x,G)
=
\partial_{G}V^\mu_\perp(x,G)-\partial_\nu V^{\nu\mu}(x,G)
=
0,
\nonumber
\ee
which is the differential equation defining the Wilson flow (gradient flow) $V^\mu_\perp(x,G)$ \cite{Luscher:2009eq}, where the flow time represents a fifth, auxiliary variable \footnote{Thanks are due to L.~Del Debbio and R.~Zwicky for encouragement to investigate if there is such a connection.}. 
Not the 5th dimensional separation $\tau$ directly, but the {worldline} propagator $G$ appears as flow-time interval. {This is due to the periodicity of the paths contributing to $w$:  
If it were not for the periodicity of the contributing paths, which in turn is due to the trace in (\ref{eq:scalar}), the equation of motion for the worldline propagator would simply be $\partial_{\tau_1}^2G(\tau_1,\tau_2)\overset{\mathrm{naive}}{=}\delta(\tau_1-\tau_2)$, which has a purely linear solution away from the delta inhomogeneity. In this case we would encounter the extra-dimensional coordinate proper in the Wilson flow. This naive worldline propagator would, however, not take into account the periodicity. For this its equation of motion needs a counter charge \cite{Strassler:1992zr} (also already because one is considering a Poisson problem on a compact interval). As a consequence the worldline propagator acquires nonlinear terms. That the worldline propagator appears in the place of the flow time means that the flow, which a priori is defined for arbitrary flow times, is probed at flow time $G$.} The $\D T$ and $\D\tau$ integrations superimpose the flow at different flow times. 
{(The above connection to the Wilson flow is not changed for fermionic quarks, as the  decisive factor $\E^{-q^2G}$ in $w_2$ remains unchanged.)}
\footnote{For the sake of 
symmetry and for facilitating the generalisation to higher correlators we can split the factor $\E^{-q^2G}=\int_0^1\D\hta\:\E^{-q^2\hta G}\E^{-q^2(1-\hta)G}$. 
Then 
\be
\tilde V_\mu^*(q)\E^{-Gq^2}\tilde V^\mu_\perp(q)
=
\int_0^1\D\hta\,\tilde V_\mu^*(q,\hta G)\tilde V^\mu_\perp[q,(1-\hta)G] .
\nonumber
\ee
}

Now, in the presence of, for instance, an attractive harmonic oscillator of strength c,
\be
\mathcal{L}
\rightarrow
\frac{\mathcal{N}_\C}{(4\pi)^2}
\int_\mathrm{P}[\D y]~\E^{-\frac{1}{4}\int_0^T\D\tau(\dot y^2-\C^2y^2+\I \dot y\cdot V)} ,
\label{eq:wlho}
\ee
where $\mathcal{N}_\C\times\int_\mathrm{P}[\D y]~\E^{-\frac{1}{4}\int_0^T\D\tau(\dot y^2-\C^2y^2)}=1$.
The potential term changes the {worldline} propagator to
\be
\C H(\tau)=\sin(\C|\tau|)-\sin(\C T)\frac{1-\cos(\C\tau)}{1-\cos(\C T)}.
\label{eq:wlprop}
\ee
Unlike $G$, $H$ is not growing monotonously without bounds for increasing $T$, which in $\E^{-q^2G}$ leads to an exponential suppression. 
The oscillations of $H$ lead 
to contributions from large $T$ that are not suppressed exponentially. 
The corresponding two-point function and flow equation are obtained by replacing $G$ by $H$.
It is this Wilson flow (gradient flow) for composite fields that extends the 4d sources to the fifth dimension of AdS$_5$. (The repulsive case is obtained for imaginary c.)

\vs

Finally, we can also analyse (\ref{eq:wlho}) in the worldline-instanton framework \cite{Affleck:1982}, for our purposes order by order in $V$. Worldline instantons are periodic classical solutions of the worldline action. The classical equations of motion possess periodic solutions only for discrete values $T=2\pi n/\C$, $n\in\mathbbm{N}$. For intermediate values of $T$ we have to determine the optimum constrained to periodic orbits. This reproduces the worldline propagator (\ref{eq:wlprop}). The periodic solutions of the classical equations of motion are marked by poles of this propagator. The determinant of the fluctuations around these classical solutions is linked \cite{Dietrich:2007vw} to the Gutzwiller trace formula \cite{Gutzwiller:1971fy}. The latter describes quantum mechanical systems (approximately, but exactly for quadratic actions) by classical quantities (periodic orbits, stability matrices, Morse indices) in analogy to the classical description of quantum field theory by holography. 

\vs

In summary, an attractive {harmonic oscillator} potential in the {worldline} formalism, a feature 
we motivated in various ways, leads to a picture that resembles AdS/QCD 
in many respects: We obtain a quantum mechanical picture involving a 5d action with hidden local symmetry over a warped AdS$_5$. The fifth coordinate is {Schwinger's proper time,} the affine parameter for the particle trajectories. The 4d sources are extended to 5d fields by a Wilson flow (gradient flow). 
More details and extensions will be presented in \cite{ddd}.

Our worldline-based approach can be extended systematically by including dynamical gluons order by order by considering higher-loop effective actions. The aforementioned step, the analysis of different gauge and chiral groups, higher correlators and other important sources, including dilaton and axion, will be discussed in later publications. (Despite the fact that the number of operators/sources grows  
exponentially with their mass dimension \cite{Cohen:2009wq}, in holography a low-energy description retaining only the lowest-dimensional operators is admissible due to the efficient decoupling of the higher operators \cite{Fitzpatrick:2013twa}.)

\vs

We expect the emergence of similar curved extradimensional formulations from the {worldline} formalism in many situations, even outside relativistic quantum field theory. The emerging spacetime will depend on the isometries of the system. For instance, Schr\"odinger (= conformal Galilean) symmetry for three spatial dimensions is reached by constraining (\ref{eq:lag}) in 4+1d Minkowski space (at $m^2=0$) by {$p^+\equiv\mathrm{const.}$} \cite{Son:2008ye}, 
which leads to a six-dimensional (i.e., with two extra dimensions) volume element $\propto T^{-7/2}$ in accordance with the construction in \cite{Son:2008ye,Balasubramanian:2008dm}.
As a related example, take a superconductor, where an attractive potential would bind the electrons into Cooper pairs. In a superconductor, however, one does not observe the tower of states, as the elementary excitation, the break-up of Cooper pairs into electrons is too far down in energy; but one can still observe the vector Cooper pairs \cite{TripletSupercon}.

\vs

The author would like to thank
S.~Brodsky,
G.~de T\'eramond,
L.~Del Debbio,
G.~Dvali,
P.~Hoyer,
M.~J\"arvinen,
J.~Reinhardt,
C.~Weiss, and
R.~Zwicky 
for  
discussions. 
The work of the author was supported by the Humboldt foundation.

\end{document}